\documentclass[aps,onecolumn]{revtex4-1}
\usepackage[dvips]{graphicx}
\usepackage{epsfig}
\usepackage{dcolumn}
\usepackage{bm}
\usepackage{amssymb}

\begin{document}
%versió 1
\title{Optimal management of impaired self-avoiding random walks for minimizing spatial coverage}

\author{Daniel Campos, Javier Crist\'{i}n and Vicen\c{c} M\'{e}ndez}
\affiliation{Grup de F\'{\i}sica Estad\'{\i}stica, Departament de F\'{\i}sica. Facultat
de Ci\`{e}ncies, Universitat Aut\`{o}noma de Barcelona, 08193 Bellaterra
(Barcelona) Spain.}

\begin{abstract}

Self-avoidance is a common mechanism to improve the efficiency of a random walker for covering a spatial domain. However, how this efficiency decreases when self-avoidance is impaired or limited by other processes has remained largely unexplored. Here we use simulations to study the case when the self-avoiding signal left by a walker both (i) saturates after successive revisits to a site, and (ii) evaporates, or dissappears, after some characteristic time. We surprisingly reveal that the mean cover time becomes minimum for intermediate values of the evaporation time, leading to the existence of a nontrivial optimum management of the self-avoiding signal. We argue that this is a consequence of complex blocking effects caused by the interplay with the signal saturation and, remarkably, we show that the optimum becomes more and more significant as the domain size increases.

\end{abstract}

\maketitle

Four decades ago Pierre Gilles De Gennes coined the suggestive expression \textit{ant in the labyrinth} to describe movement through disordered systems \cite{gennes76}. It is widely known that by introducing obstacles in regular lattices the effective diffusion coefficient of random walkers gets reduced proportionally, and eventually transport becomes subdiffusive when the percolation threshold is reached due to the self-similar properties of the underlying structure \cite{bunde91}. Despite the intrisic complexity of the problem, throughout the years effective propagators and Fokker-Planck equations have been proposed, and its main scaling properties have been progressively revealed \cite{oshaugnessy85,metzler94,drager95,drager96,avraham00,campos04,bianco13,balankin15}. 

A far less understood situation, however, is that in which disorder is not quenched but dynamically generated by the trajectory itself. Some well-known models fulfilling this idea are self-avoiding or self-repelling random walks, in which revisits to previous nodes/positions are systematically avoided throughout the trajectory. So, strong non-Markovian effects govern the dynamics of these systems, which turns their analytical treatment cumbersome in most cases. Nevertheless, the interest of self-avoiding walks as stochastic processes for optimizing exploration or coverage of the media is evident, as they represent a way to consistently avoid overlaps typical of recurrent trajectories (e.g. Brownian paths), specially in low-dimensional systems.

Coverage optimization through self-avoiding rules is potentially attractive for sampling efficiently large phase spaces (for instance, for Monte Carlo algorithms in statistical mechanics) \cite{nemirovski90}. Furthermore, the concept of self-avoidance is also important to understand the dynamics of particles which are able to leave locally some kind of signal or debris which can yield a local repulsive potential afterwards \cite{grassberger17}. Such systems are gaining nowadays a renewed interest due to the growing experimental evidence that many microorganisms like bacteria or T-cells could be able to use self-signalling mechanisms for increasing their dispersal, feeding, or predation efficiencies \cite{tweedy16,dona13,schwab07,tweedy16b}, and also due to the availability of new techniques for generating controllable artificial self-repelling particles in the lab, e.g. microdroplets in surfactant solutions \cite{jin17,liebchen18}. Finally, self-avoidance can be seen as a mechanism for optimizing searches, for instance during animal exploration/foraging \cite{berbert12,sims14,reynolds14,mendez14,sakiyama18} or in search algorithms through the Internet \cite{avin07,millan12,oshima12,kim16,arruda17,herrero19} or in social networks \cite{gong13,bagnato18}, among other.

A reference model within this context is the true Self-Avoiding Walk (tSAW), first introduced by Amit, Parisi and Peliti \cite{amit83} as a way to disentangle self-avoiding random walks from models of polymer growth, as the latter are known to be typically self-killing instead of self-avoiding \cite{grassberger17}. The tSAW rule of advance work as follows: given a present position of the walker, the probability to jump in the next step to each of the first neighbours $j$ is $p_{j}=Z^{-1} e^{-g n_j}$, with $Z \equiv \sum_{j=1}^{z} e^{-g n_j}$ a normalization factor where $z$ is the coordination umber of the lattice, $g$ a positive constant and $n_j$ (denoted here as the signal intensity) is the number of visits that the walker has made to node $j$ previously. Accordingly, those neighbours less visited in the past are preferentially selected, with $g$ controlling the prevalence of the self-avoidance. 

Coverage properties of classical random walks moving within regular (finite) lattices in $d$ dimensions have been extensively explored over the last thirty years \cite{yokoi90,nemirovski90,aldous91,nemirovski91,brummelhuis92,freund93,coutinho94,dembo04,grassberger17b,cheng18}. The coverage problem in $d=1$, for example, can be mapped to a first-passage problem and then analytical expressions can be obtained for the mean time required to cover all nodes in the domain, $\langle T_{cov} \rangle = N (N-1)$, with $N$ denoting the number of nodes in the lattice \cite{yokoi90}. Also, the case $d=2$ has been proved to satisfy $\langle T_{cov} \rangle \sim N (\log{N})^2$, while for $d \geq 3$ it is found that $\langle T_{cov} \rangle \sim N (\log{N})$ \cite{brummelhuis92,grassberger17b}. Furthermore, universal scaling properties have been revealed recently to emerge in the distribution of coverage times for non-recurrent random walks in different dimensions \cite{chupeau15}. An equivalent analysis for the tSAW, on its turn, becomes more complicated due to the memory effects involved. Still, we know that for $d=1$ the case of perfect self-avoidance ($g \rightarrow \infty$) will result in perfect coverage (i.e. $\langle T_{cov} \rangle \sim N$), while for low values of $g$ the scaling $\langle T_{cov} \rangle = N (N-1) \sim N^2$ of regular walks should be recovered. For $d=2$ the scaling $\langle T_{cov} \rangle \sim N (\log{N})$ for large $g$ has been conjectured in \cite{avin07} and confirmed numerically in \cite{grassberger17}. Also, since the critical dimension of the tSAW is known to be $d=2$ \cite{amit83}, the scaling is expected to be identical to that of regular random walks for $d > 2$. 

Despite all this findings, there are very few works in the literature that have explored how the properties of thse models get modified when self-avoidance is limited and/or impaired (see \cite{moreau09, grassberger17} as some interesting exceptions). If we consider the potential applications mentioned above (e.g. in self-repelling trajectories of microdroplets or microorganisms resulting from chemical signals) it is natural to wonder about the effects that diffusion or evaporation (among other) of these signals will have on the properties of the corresponding trajectories, and on their coverage efficiency. This idea has been addressed recently for a modified version of the tSAW introducing dispersal of the chemical through a variation of the signal levels in the neighbouring nodes whenever a site is visited; this model has led to the surprising observation that tSAWs can become self-trapping in some situations \cite{grassberger17}. Our present work offers an alternative view within this context, by exloring how tSAW coverage properties are modified if (i) the effect of the self-avoiding signal is assumed to become less and less effective as long as successive visits to a node are performed (we call this \textit{signal saturation}), and (ii) the signal can dissappear with time (we call this \textit{signal evaporation}). As we will show, several unexpected facts arise as a consequence of these restrictions. In particular we observe that increasing the rate of \textit{evaporation} does not always result in a larger coverage time, but an optimal \textit{evaporation} time can exist for systems below the critical dimension of the tSAW. This effect, as we shall see, is modulated by the intensity of the \textit{signal saturation}, and its significance increases as long as the size of the system grows. 

First of all, we implement the idea of \textit{signal saturation} by considering that the signal intensity is of the form $n_j= \sum_i i^{-\gamma}$ (with $\gamma$ a positive parameter), where the sum is carried out over all the previous visits of the walker to that site. Then, the first visit to the site ($i=1$) will increase the intensity in one unit while the increase will be smaller for subsequent visits. In particular, for $\gamma =0$ we recover the classical rule of the tSAW, while in the limit $\gamma \rightarrow \infty$ the walker is only able to distinguish visited from nonvisited sites (but it cannot distinguish, or \textit{remember}, how many times the site has been revisited).

The dependence of $\langle T_{cov} \rangle $ on $N$ obtained for this model as a function of the $\gamma$ parameter is presented in Fig. 1. We can observe that for the classical tSAW ($\gamma =0$) and $g$ large, the scaling is approximately $\langle T_{cov} \rangle \sim N$ (for $d=1$) and $\langle T_{cov} \rangle \sim N \log{N}$ for $d=2$, in agreement with the results in \cite{grassberger17}. 

For larger values of $\gamma$ one should note that intuitively \textit{signal saturation} plays a similar role to that of decreasing $g$ with time, fading gradually the self-avoiding mechanism. In consequence nee should expect a behavior
\begin{eqnarray}
\nonumber \langle T_{cov} \rangle \sim N^{\alpha} \quad (d=1) \\
\nonumber \langle T_{cov} \rangle \sim N \left( \log{N} \right) ^{\alpha} \quad (d=2)
\label{eq1}
\end{eqnarray}
with $\alpha$ increasing gradually from the tSAW value ($\alpha =1$) to the value of the regular random walk ($\alpha=2$). This is approximately what we find from our simulations, except that the value of $\alpha$ is often found to take values larger than $2$. This means that there are some regions of parameters where the coverage time increases even faster with $N$ that it would for a regular random walk. This already gives us the idea that \textit{signal saturation} can reduce drastically the efficiency of the coverage. Appart from that, the role of \textit{signal saturation} in the tSAW framework is relatively trivial \textit{per se}, since it is rather equivalent to reducing self-avoiding accuracy (i.e. decreasing $g$ gradually). 
%It is also worth mentioning that in all the cases, above the critical dimension of the tSAW the scaling is not modified at all by these effects; so, for $d=3$ we have checked that $\langle T_{cov} \rangle \sim N \log{N}$ emerges always independently of $\gamma$ and $g$ (not shown here).

The situation becomes far less trivial when \textit{signal evaporation} is taken into account. This should be implemented by decreasing the self-avoiding signal intensity $n_j$ at every site $j$ at a given rate. However, in order to simplify computational work we consider here an all-or-nothing rule in which a random time $t_j$ (according to an exponential probability distribution function, $\rho(t _j)= \tau^{-1} e^{-t_j / \tau}$) is chosen whenever the walker visits a given site $j$, and the signal intensity $n_j$ at that node is reset to zero at a time $t_j$ after the visit. The parameter $\tau$ then represents the characteristic timescale at which the memory of the signal is completely lost by the effect of \textit{evaporation}. While all the results reported in the following have been obtained through this all-or-nothing rule, our numerical analysis have revealed that our conclusions would remain qualitatively the same if a progressive \textit{evaporation} rate $\sim \tau^{-1}$ was considered instead.

For the case with both \textit{saturation} and \textit{evaporation}, we find that there is no simple general scaling of $\langle T_{cov} \rangle$ with $N$ except in the trivial limit $\tau \gg \langle T_{cov} \rangle$ for which the results above are recovered; as $\tau$ is reduced the exponent $\alpha$ introduced in (\ref{eq1}) seems to depend on $N$ itself, and it progressively increases as $N$ grows. But the most surprising fact occurs when the mean cover time is computed as a function of the evaporation time $\tau$ (Fig. 2). While $\langle T_{cov} \rangle$ decreases monotonically with $\tau$ when $g$ is large and $\gamma$ is low, an optimal evaporation time emerges for moderate values of these two parameters. In Fig. 2 one can see how this effect depends on the value of $\gamma$. As a whole, an optimum $\tau_{opt}$ is found whenever the self-avoiding mechanism is not accurate enough because of a low $g$ and the presence of the \textit{signal saturation}. The combination of both seems to be necessary since the optimum does not appear for $\gamma =0$ (but see our comments below). Also, this phenomenon is found to be characteristic of low-dimensional lattices, but it dissappears for systems above the critical dimension of the tSAW.

Figure 3 provides a deeper insight by showing how the optimum evaporation time, $\tau_{opt}$ depends explicitly on $\gamma$ and $g$ for both $d=1$ and $d=2$. It is clear there that, if $\gamma$ is large enough, a transition always occurs at a given value of $g$ above which the optimum coverage is simply $\tau \rightarrow \infty$, while for lower values of $g$ the nontrivial optimum emerges. Since that critical value seems to depend explicitly on $\gamma$ and $N$, we cannot interpret this in terms of a classical phase transition (we have actually checked that the behavior around the critical value does not reproduce the scaling properties of critical points in thermodynamic systems). Still, it is clear that there is a qualitative difference in the dynamics of coverage for low and high values of $g$.

To understand why this occurs, we must take into account that \textit{signal saturation} makes self-avoidance effects to become uneffective after some time, by decreasing progressively the signal gradients. So, it drives the system towards a more or less uniform landscape (it is, $n_i - n_j \rightarrow 0$ for any pair of neighbour nodes $i$ and $j$). In that limit, the tSAW will eventually behave as a regular random walk, and so it will be extremely unefficient at covering the domain. This effect will be specially relevant if $g$ is small so the coverage process takes long times to complete. For $g \rightarrow \infty$, instead, the limit of a uniform energy landscape will be only reached at a timescale much longer than $\langle T_{cov} \rangle$ and so this will not have any effect on the coverage. 

Within this context, paradoxically, \textit{signal evaporation} becomes benefitial since it continually offers new 'seemingly unexplored' regions to the walker. The walker will tend to go to that regions and this will enhance its global mobility, so avoiding (at least partially) recurrent paths characteristic of a regular random walk. This can be visualized in Fig. \ref{fig5}, where we show a particular realization of the landscape of $n_j$ for a $32 \times 32$ lattice at the time of completing the coverage process, and for different values of $\tau$. Values close to the optimum one (middle panel) correspond to a complex mix of visited ($n_j \geq 1$) and forgotten ($n_j=0$) sites, if compared to the case of a much lower (left) or higher (right) $\tau$ value.

As a whole, our analysis shows that whenever the self-avoiding mechanism has a vanishing accuracy, then \textit{signal evaporation} (or, equivalently, forgetting part of the previous path) is actually benefitial in terms of  coverage, since it avoids getting 'trapped' or 'blocked'. We can interpret the path of the walker then as switching between two different transport regimes: (i) time intervals in which the walker finds a new region to explore (either because the region is really unvisited, or because signal has \textit{evaporated} from there) and then self-avoidance can efficiently act, and (ii) time intervals in which the walker gets trapped into a region of homogenous energy and transport becomes diffusive. Escape from these regions of regime (ii) could then possibly understood as a Brownian escape problem from a region whose size and shape changes dynamically with time. Optimal coverage strategies then result from the balance between promoting escape by enhancing \textit{evaporation} at some intermediate levels.

Regarding the dependence on the system size $N$, as a first approximation it would be reasonable to expect that $\tau_{opt}$ will be proportional to $N$. However, as $N$ increases larger and larger 'trapping' regions will emerge and their effect will become proportionally larger. To verify this, we show in Fig. \ref{fig4} how the shape of the 'blocked phase' changes as a function of $N$. Here we denote 'blocked phase' as the region of parameters for which a finite $\tau_{opt}$ appears, while the 'normal phase' corresponds to the standard situation with $\tau_{opt} \rightarrow \infty$. In agreement with our previous discussion, the size of the 'blocked phase' in Figure \ref{fig4} grows monotonically with $N$; this comes together with a decrease of the relative optimum $\tau_{opt}/N$ for fixed values of $\gamma$ and $g$ (not shown). Eventually, this would lead in the thermodynamic limit $N \rightarrow \infty$ to the remarkable fact that the optimum \textit{evaporation} time becomes vanishingly small, i.e. $\lim_{N \rightarrow \infty} \tau_{opt} / N = 0$ and so $\lim_{N \rightarrow \infty} \tau_{opt} / \langle T_{cov} \rangle = 0$.

Finally, we mention for the sake of completeness that, albeit all the results presented here correspond to lattices with periodic boundary conditions, the effect of considering reflecting boundaries, for example, will clearly enhance the 'blocking' effect (through the possibility of getting 'trapped' near the boundaries), and so it will increase the range of parameters for which a finite $\tau_{opt}$ exists. As a proof of concept, we have checked that, at least in $d=1$, the optimum $\tau_{opt}$ can appear for reflective boundary conditions even in the absence of \textit{signal saturation} (this is, when $\gamma =0$); additional details and discussions will be presented in a forthcoming publication.

In summary, we have presented here a previously unreported phenomena about the coverage properties of self-avoiding random walks. While self-avoidance is typically assumed to represent an extremely efficient mechanism for domain coverage, we have proved numerically that a vanishing accuracy in self-avoiding yields the existence of an optimal memory management. This means that under certain circumstances it becomes more efficient to \textit{forget} part of the regions previously covered (through an \textit{evaporation} process) that keeping full memory of the path. This has been illustrated for the paradigmatic case of the tSAW, but we claim that the existence of the 'blocked' phase (and so of an optimal memory management) will presumably appear in many other self-avoiding models, as well as in alternative random walk models with memory under similar conditions. A deeper analysis of this phenomena, then, can open a useful line of research in order to promote our understanding about how to optimize the efficiency of artificial self-repelling microparticles (for medical applications, for instance) or self-avoiding searches on networks, or about the evolutionary forces that may had driven the improvement of cognition and memory for navigation and/or foraging in living beings.

\newpage

\begin{figure}
	\includegraphics[scale=0.9]{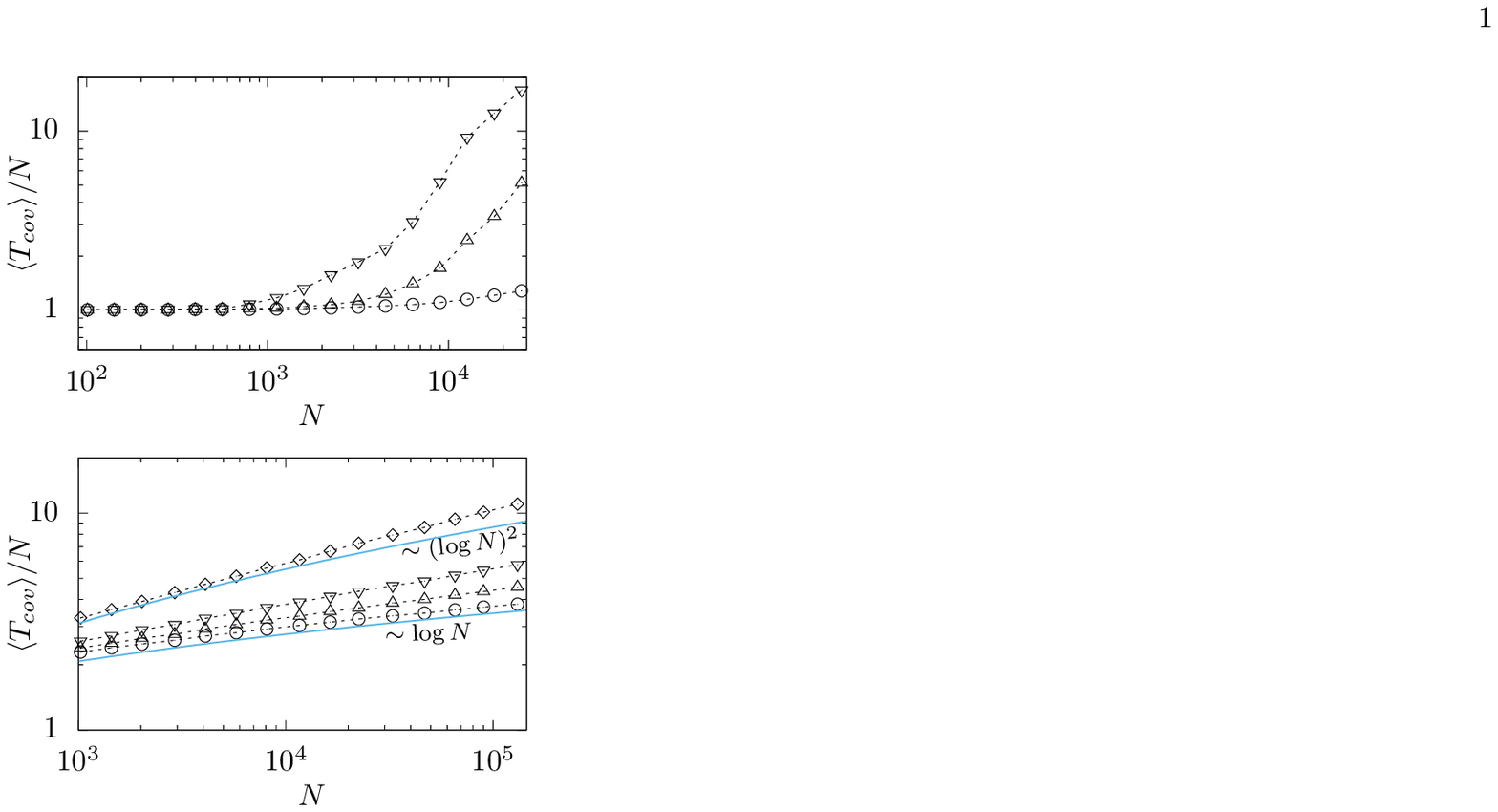}
	\caption{Scaling of the mean time coverage $\langle T_{cov} \rangle$ for regular lattices in $d=1$ (upper plot) and $d=2$ (lower plot) for different intensities of the \textit{signal saturation} parameter $\gamma$, with $g=10$. Symbols correspond to values obtained from simulations: circles ($\gamma=0$), triangles ($\gamma=1$), inverted triangles ($\gamma=2$) and diamonds ($\gamma=4$), while dotted lines are just to facilitate visualization. Solid lines show logarithmic and square logarithmic scalings for comparison purposes.}
	\label{fig1}
\end{figure}

\begin{figure}
	\includegraphics[scale=0.9]{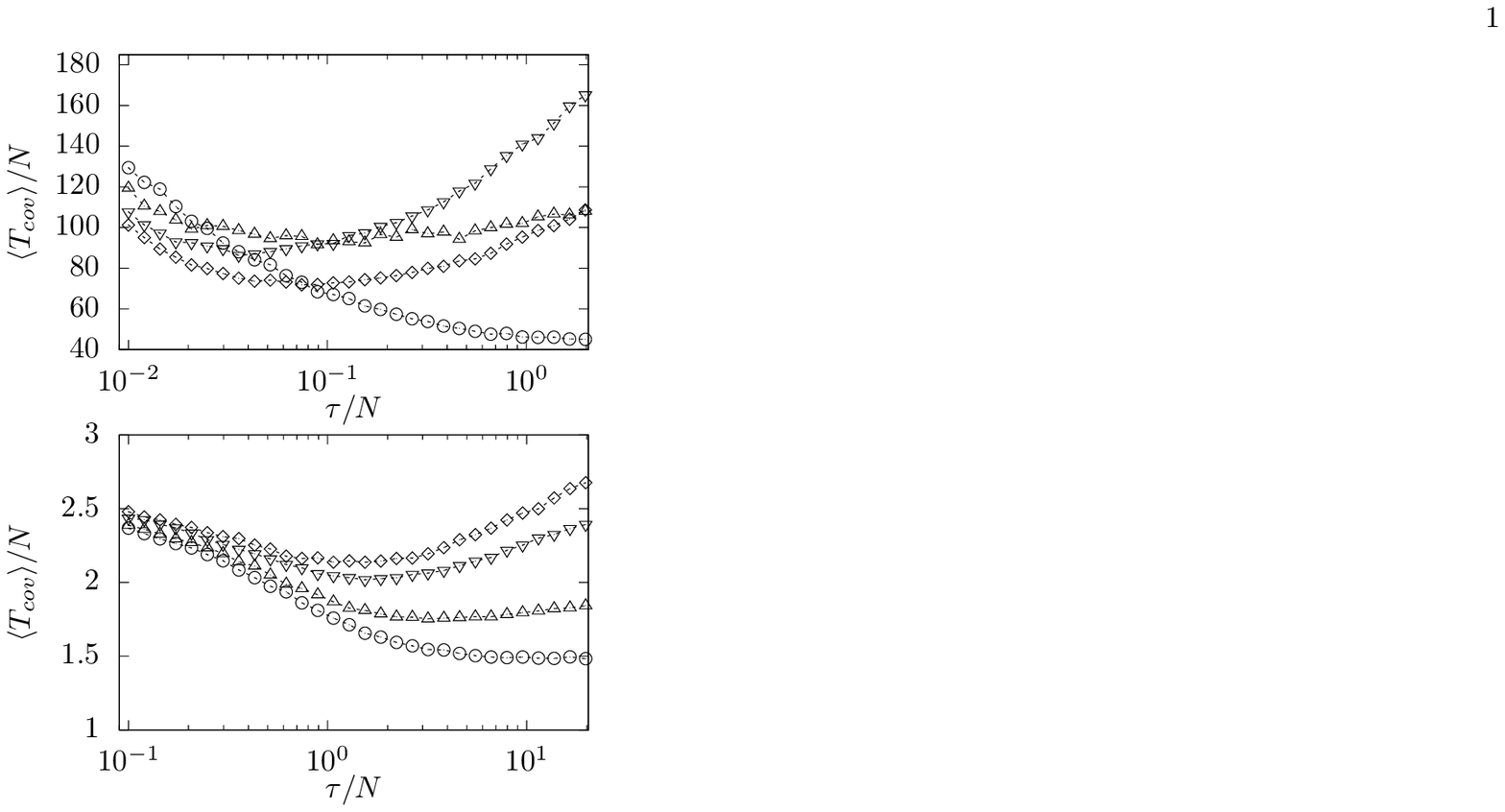}
	\caption{Mean time coverage as a function of the characteristic \textit{evaporation} time. For $d=1$ (upper plot) we take $g =2$ and $N=5 \times 10^{3}$, and show plots for $\gamma =0$ (circles), $\gamma =1$ (triangles), $\gamma =2$ (inverted triangles) and $\gamma =3$. (diamonds). For $d=2$ (lower plot) we take $g =10$ and $N=128 \times 128$, and show plots for $\gamma =1$ (circles), $\gamma =2$ (triangles), $\gamma =3$ (inverted triangles) and $\gamma =4$. (diamonds)}
	\label{fig2}
\end{figure}

\begin{figure}
	\includegraphics[scale=0.9]{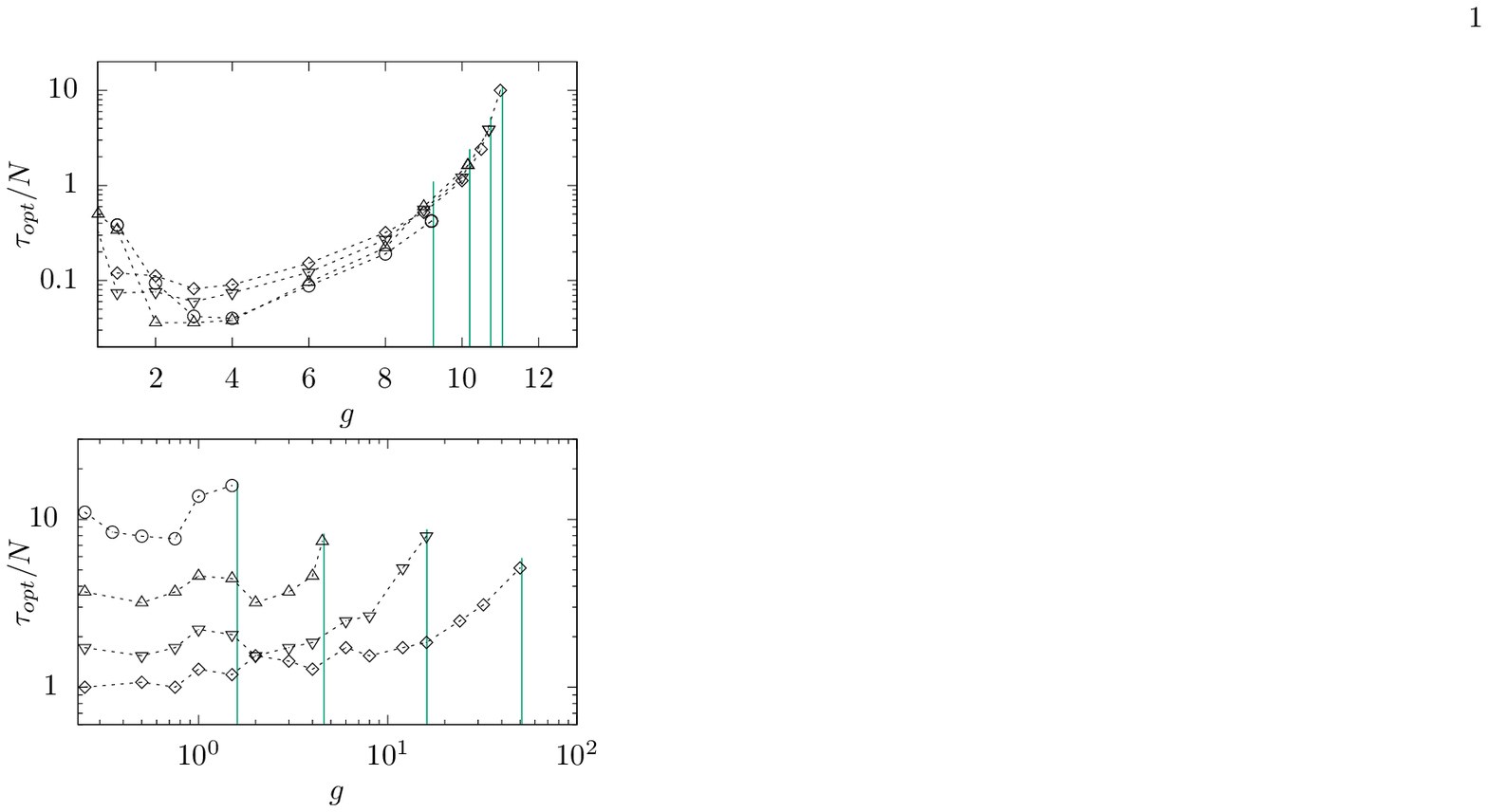}
	\caption{Optimum \textit{evaporation} time as a function of $g$. Results for $d=1$ (upper plot) correspond to $N=5 \times 10^{3}$ and $\gamma =1$ (circles), $\gamma =2$ (triangles), $\gamma =3$ (inverted triangles) and $\gamma =4$. (diamonds). Results for $d=2$ (lower plot), on its turn, correspond to $N=128 \times 128$ and $\gamma =1.5$ (circles), $\gamma =2$ (triangles), $\gamma =3$ (inverted triangles) and $\gamma =4$. (diamonds). Vertical lines indicate the $g$ threshold (for each $\gamma$) above which a finite optimum $\tau _{opt}$ does not exist.}
	\label{fig3}
\end{figure}

\begin{figure}
	\includegraphics[scale=0.9]{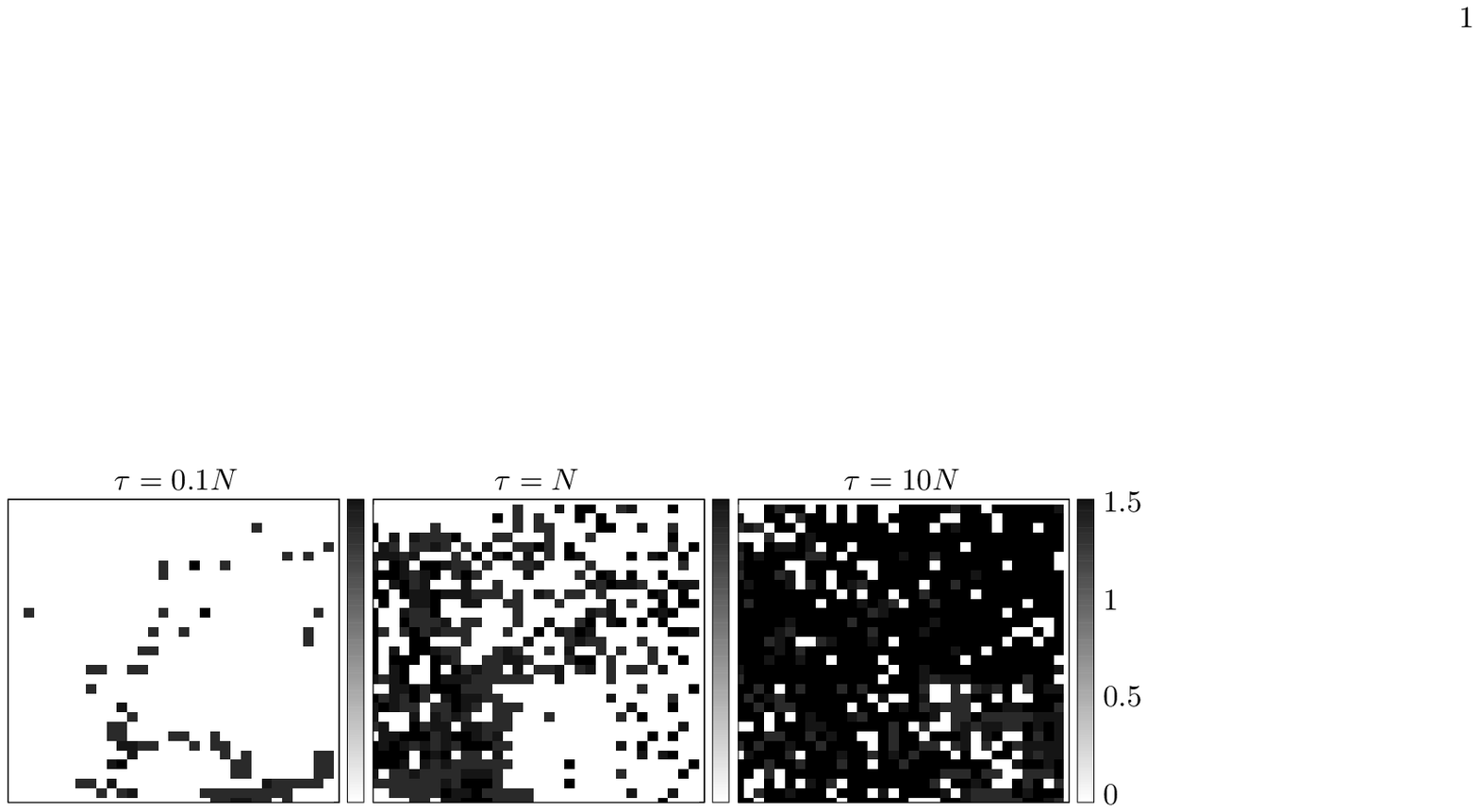}
	\caption{Signal intensity landscape for a situation close to the optimum $\tau_{opt}$ (case $\tau=N$) compared to much lower (left panel) and higher (right panel) values of $\tau$ (see labels). The maps shown correspond to a particular realization of our self-avoiding model evaluated at the coverage time $t=T_{cov}$. The gray colors in the plot represent the values of $n_j$ for each node $i$ in a $32 \times 32$ lattice, according to the legend on the right. The values of the parameter $g =2$ and $\gamma =3$ have been used in all cases.}
	\label{fig5}
\end{figure}

\begin{figure}
	\includegraphics[scale=0.9]{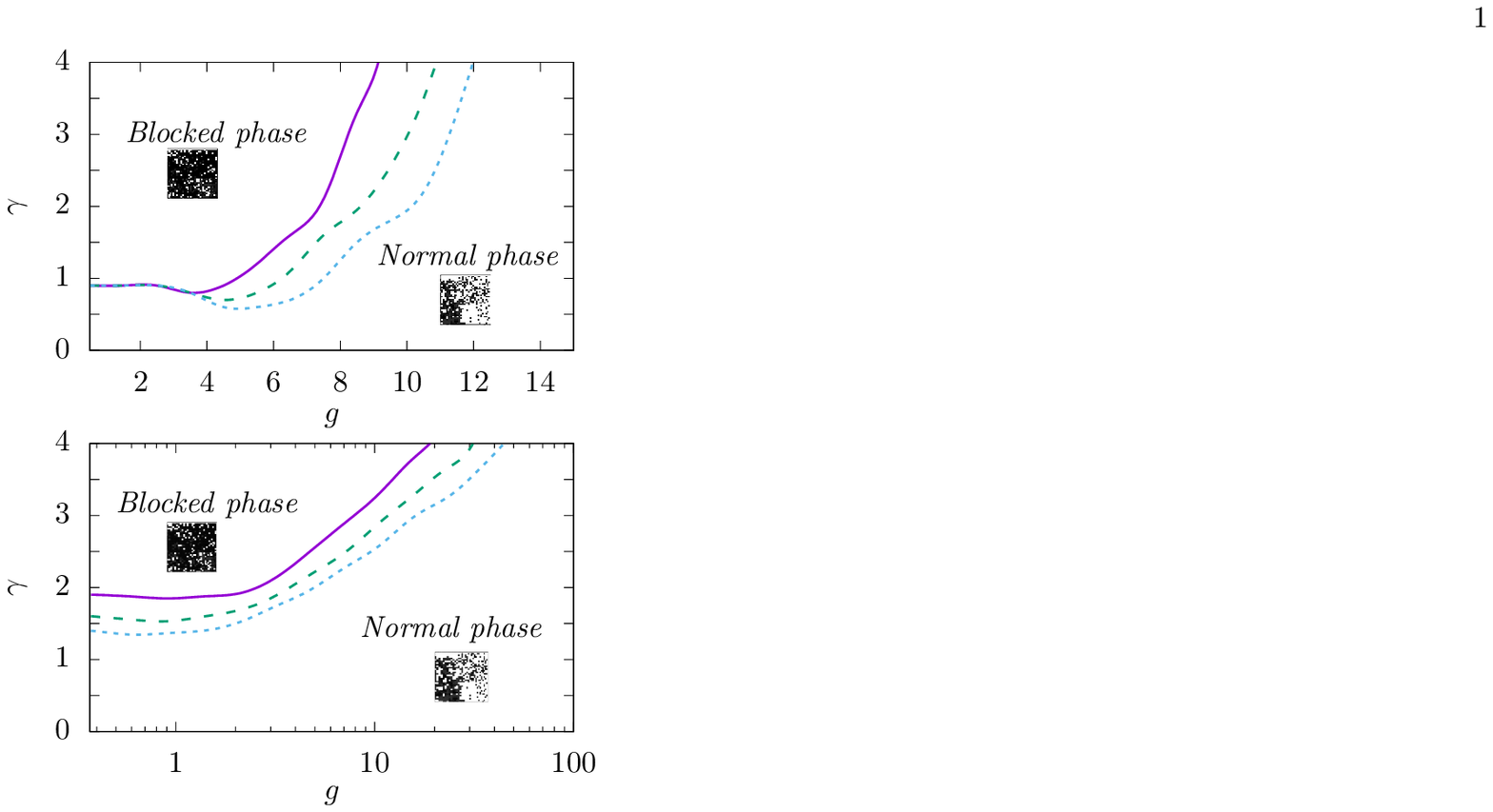}
	\caption{Phase diagram $\gamma$-$g$ separating the 'blocked' phase, for which a finite $\tau _{opt}$ exists, from the normal region where it does not. The different lines correspond to different values of $N$. For $d=1$ (upper plot) the lines correspond to $N=10^{3}$ (solid), $N=2 \times 10^{3}$ (dashed) and $N=5 \times 10^{3}$ (dotted). For $d=2$ (lower plot), $N=32 \times 32$ (solid), $N=64 \times 64$ (dashed) and $N=128 \times 128$ (dotted) have been used.}
	\label{fig4}
\end{figure}

\end{document}